\documentclass{epl}
\usepackage{amssymb}

\title{Chaotic quantum decay in driven biased optical lattices}
\shortauthor{S. Mossmann \etal}
\author{S. Mossmann\inst{1,2} \and C. Schumann\inst{1} \and  
H. J. Korsch\inst{1}\thanks{E-mail: \email{korsch@physik.uni-kl.de}}}
\institute{
  \inst{1} Fachbereich Physik, Technische  Universit\"at Kaiserslautern, D-67653 
              Kaiserslautern, Germany\\
  \inst{2} present address: Centro de Ciencias F\'isicas, UNAM, M\'exico
}

\begin{document}

\maketitle

\begin{abstract}
Quantum decay in an ac driven biased periodic potential modeling
cold atoms in optical lattices is studied for a symmetry broken
driving. For the case of fully chaotic classical dynamics
the classical exponential decay is quantum mechanically
suppressed for a driving frequency
$\omega$ in resonance with the Bloch frequency $\omega_B$,
$q\omega=r\omega_B$ with integers $q$ and $r$.
Asymptotically an algebraic decay  $\sim t^{-\gamma}$ is observed.
For $r=1$ the exponent $\gamma$ agrees with $q$ as
predicted by non-Hermitian random matrix theory for $q$
decay channels. The time dependence of the
survival probability can be well described by random matrix theory. 
The frequency dependence of the survival probability shows
pronounced resonance peaks with sub-Fourier character.
\end{abstract}

\section{Introduction} Chaotic classical dynamics
has quantum signatures in statistical properties of eigenvalues,
wave amplitudes and state projections \cite{Haak01} which can be
described by random matrix theory. This has been 
demonstrated in many cases for bound systems. For open systems, 
however, such studies
are rare. Only recently, the statistics of the lifetimes 
of the metastable resonance states (or the
complex part $\Gamma$ of the resonance energies) have been shown 
to be in agreement with the appropriate random matrix results. 
This has been achieved for only few physical
systems, e.g.~open billiards \cite{Ishi95,Wirt97,Ishi00},
scattering on graphs \cite{Kott97Kott00} and strongly ac driven 
periodic lattices
\cite{99delay99lifetime,02wsrep} which will also be studied in the 
present letter.

Open classical systems with fully chaotic intrinsic dynamics show
generically an exponential decay of the survival probability
\,$P(t) = e^{-\nu t}$\, in the long time limit, where the decay 
rate $\nu$ can be related to
the Lyapunov exponent and the fractal dimension of the chaotic repeller
\cite{Gasp98}.
Quantum mechanically, the situation is more complicated because
the number of open decay channels depends on the
Planck constant $\hbar$ and typically increases in the limit of
small $\hbar$. 

\section{The model}
The one-dimensional Hamiltonian
\begin{equation}
\label{hamiltonian}
H=p^2/2+V(x)+F(t)x \ ,\quad V(x+2\pi)=V(x) \;,
\end{equation}
with space period $d=2\pi$ and  ac-dc driving
\begin{equation}
F(t)=F_0+F_\omega(t)\ ,\quad F_\omega(t+T_\omega)=F_\omega(t)
\end{equation}
(where the time-average of $F_\omega(t)$  can be chosen to be zero)
is known as the ac-dc Wannier-Stark system.
Such systems have been studied
by many authors in different context (see, e.g.~the  review \cite{02wsrep}). 
In recent years, one
could observe an increasing interest in view of the exploding
studies on the dynamics of cold atomic gases or Bose-Einstein
condensates in optical lattices, where the static field $F_0$ is
generated by chirping the laser frequencies or simply by the gravitational
force.

Note that in the scaled units used here, $\hbar$
depends on the system parameters and can be expressed as 
$\hbar = 4\sqrt{E_R/V_0}$ where $V_0$ is the potential depth 
and $E_R = \hbar^2k^2/(2m)$ is the recoil energy  \cite{02wsrep}.
In present experiments a ratio of $V_0/E_R\approx 500$ is 
routinely available which yields $\hbar \approx 0.2$
not far from the value $\hbar =0.1$
used in the numerical calculations below.

It is convenient to rewrite the dynamics in a Kramers-Henneberger
form \cite{02wsrep}
\begin{equation}
\label{eq:KHHamiltonian}
H_{\rm KH}=(p-F_0t)^2/2+V(x-K_\omega(t)\,)
\end{equation}
with
$K_\omega(t)=\int_{t_0}^{t}\!\!d t'\,G_\omega(t')
\,,\ G_\omega(t)=\int_{t_0}^{t}\!\!  d t'\,F_\omega(t')\,.$
We will study the potential $V(x)=\cos x$ with de-symmetrized driving
\begin{equation}
F_\omega(t)=A_\omega\left(\,\cos\omega t+\sin 2\omega t\,\right)
\label{Fomega}
\end{equation}
and we choose the time $t_0$ as a solution of 
\,$2\sin \omega t_0= \cos 2\omega t_0$\, with the consequence
that  
\begin{equation}
K_\omega(t)=-\frac{\epsilon}{4}\,\big(4\cos \omega t+\sin 2\omega t
-4\cos \omega t_0-\sin 2\omega t_0\big)
\end{equation}
($\epsilon =A_\omega/\omega^2$) is periodic.

\section{Chaotic dynamics}
\begin{figure}[t]
\twoimages[width=7cm, clip]{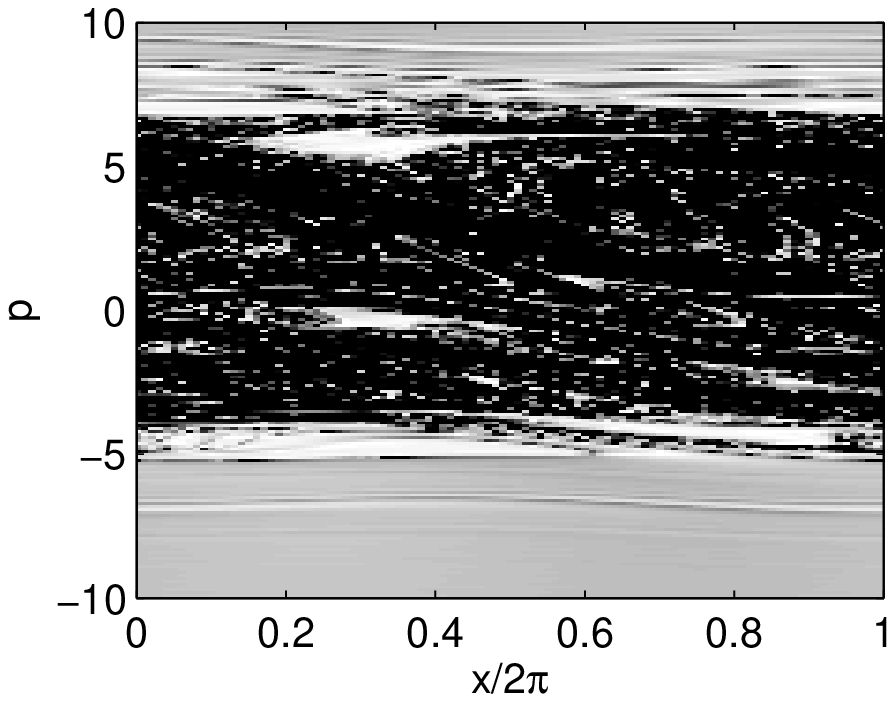}{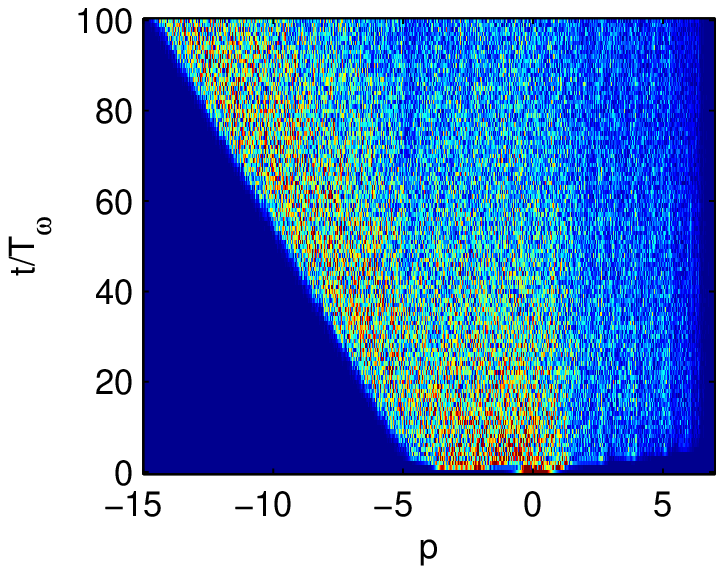}
\caption{Norm of the monodromy matrix as a function of the
initial phase space point $(x,p)$ for a time propagation
over six periods (left) and time evolution of a wavefunction $|\psi(p,t)|$ 
initially localized in the chaotic strip (right). The parameters are
$\omega=1$, $\epsilon=3$ 
with a static field is $F_0=0.016$.}
\label{phsp-wave}
\end{figure}

Let us first consider the classical dynamics. A stroboscopic 
phase space plot for the non-biased case  $F_0=0$, e.g.~for parameters
$\omega=1$, $\epsilon=3$,
reveals a chaotic strip 
with few embedded
regular islands. When a weak static force $F_0$ is added, this chaotic strip
survives but the invariant curves confining the chaotic strip are destroyed.
A trajectory starting inside the strip will show a diffusive motion
until it reaches the (former) boundary $p_1$ where it escapes to infinity.
As a measure of the local spreading in phase space we calculate
the stability (or monodromy) matrix $M(t)$ \footnote{For a bound
system, the long time average of the norm of the stability matrix
gives the Lyapunov exponent. Here the majority
of the trajectories escape to infinity and this limit is not defined.
For more information on phase space delocalization see \cite{02deloc}.}
for a trajectory 
started at a phase space point $(x,p)$.
Figure \ref{phsp-wave} shows the norm $||M(t)||$ at time
$t=6T_\omega$ as a function of $(x,p)$ for
$\omega=1$, $\epsilon=3$
and an additional static force $F_0=0.016$, a value used
also in the quantum case discussed below for resonant driving.
We clearly observe the chaotic strip where the lower boundary at 
$p_1\approx -4$ is relatively sharp whereas the upper boundary
at $p_2\approx +6$
is more diffusive. This difference is due to the fact that particles
with large enough negative momentum escape and particles with
positive momentum are reflected by the increasing linear potential
and re-injected into the chaotic strip. 
A similar behavior is found for other values of $F_0$
used in the quantum calculations discussed in the following.

The quantum dynamics is qualitatively different because
of the Bloch oscillation \cite{04bloch1d}, an intrinsic periodic motion with the Bloch 
period in the non-driven case $F_\omega(t)=0$.
By constructing a generalized scattering-matrix for such systems, resonance
states can be defined and also easily computed \cite{02wsrep}, provided
that the driving and the Bloch frequency 
$\omega_B=d F_0/\hbar$ are in resonance,
\begin{equation}
q\,\omega= r\,\omega_B\,, 
\label{q-r}
\end{equation}
where $r$ and $q$ are coprime integers. The integer $q$ is equal to
the dimension of the scattering matrix, i.e.~the number of open channels
\cite{02wsrep}. 
This offers the unique possibility to tune the
number of decay channels by varying the external system parameters.

Here we will explore the quantum decay dynamics for $r=1$ as a function of the
number $q$ of decay channels, starting from the case of resonant driving,  
$\omega=\omega_B$, i.e. $q=1$. Here and in the following we use $\hbar=0.1$. 
The quantum wavefunction $|\psi(p,t)|$
is plotted in fig.~\ref{phsp-wave} for an initial minimum
uncertainty Gaussian wavepacket  with momentum width $0.28$ centered at $p=0$ 
in the classically chaotic region.
Also here the chaotic strip can be easily identified:
The wavefunction rapidly spreads over the strip and decays in the direction
of negative momentum. Inside the strip we find an irregular oscillation
and outside, for $p<p_1$, a decay to $-\infty$ with linearly growing 
momentum $p\sim -F_0t$. If the assumption of an 
irregular wavefunction inside the chaotic strip is correct, the statistics
of the normalized probabilities
$s_n=|\psi(p_n,t)|^2/\sum_n |\psi(p_n,t)|^2$ at discrete values of the
momentum should follow the random vector model 
\cite{Haak01,Zimm89,98ent} 
describing in general the fluctuations of a sum of $\nu$ independent 
Gaussian distributed variables ($\nu=1,2,4$) for the
orthogonal (GOE), unitary (GUE) and symplectic (GSE) ensemble, respectively. 
This model predicts approximately a $\chi^2_\nu$ distribution 
\begin{eqnarray}
\label{chi-dist}
W_\nu(s)\approx \chi_\nu^2(s) = \left(\frac{\nu}{2}\right)^{\nu/2}\frac{s^{\,\nu/2-1}}{\Gamma(\nu/2)}
\,{\rm e}^{-\nu s/2}\,.
\label{random-nu}
\end{eqnarray}

\begin{figure}[t]
\onefigure[width=7.5cm, clip]{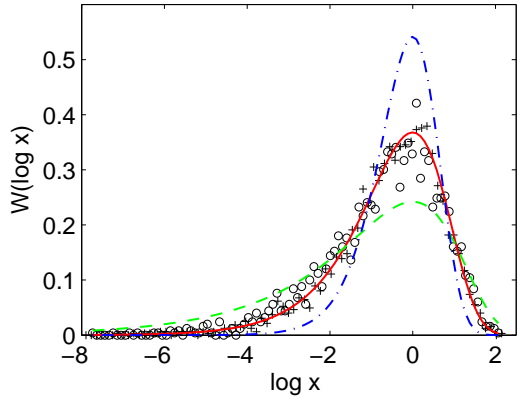}
\caption{Statistical distribution of $s=|\psi(p,100\,T_\omega)|^2$ within the 
chaotic strip for $q\!=\!1$ ($+$) and $q\!=\!2$ ($\circ$) in comparison 
with the random matrix predictions (\ref{random-nu}) for the GOE
($- \cdot - \cdot$), GUE (---) and the GSE (- - -) case. 
Shown is the distribution $W(\log x )$ with $x= s/\bar s$
in order to increase the differences\,.}
\label{wavestat}
\end{figure}

Figure \ref{wavestat} shows a  comparison of the numerical data
of fig.~\ref{phsp-wave} in the chaotic strip for $t = 100\,T_\omega$
with the distributions (\ref{random-nu}) for $\nu =1,2,4$. 
As expected, we observe an agreement with the GUE 
statistics. Moreover, this behavior is found to be independent of the ratio
$\omega/\omega_B=1/q$. The decay, however, is very sensitive with respect
to this ratio. 

\section{Dynamics of decay}
In the following, we will explore the validity of the conjecture that the
decay of these Wannier-Stark systems can be described by
non-Hermitian random matrix theory for systems with
$q$ decay channels.
 
According to random matrix theory,
the distributions of the resonance widths $\Gamma$ 
of the Wannier-Stark resonance states (the eigenstates of the Floquet operator) 
are given by \cite{Fyod96}
\begin{eqnarray}
\label{gamma-dist}
\Pi(\Gamma_s) =  \frac{(-1)^q}{(q-1)!} \,\Gamma_s^{q-1} 
\frac{d^{\,q}}{d \Gamma_s^{\,q}}\left[ 
\frac{1 - \exp(- 2 \Gamma_s)}{2 \Gamma_s} \right]\, , 
\end{eqnarray}
for the circular unitary ensemble (CUE). 
Here $q$ is the number of channels and
$\Gamma_s=\pi \Gamma/\Delta$ is the width scaled by the mean level 
spacing $\Delta$.
Equation (\ref{gamma-dist}) can also be aplied to the case of a harmonic driving as
shown in~\cite{02wsrep} if one restricts the Hamiltonian to a fixed value of the
quasimomentum $\kappa$ which is a good quantum number. With the exception
of the center and the boundaries of the Brillouin zone the time reversal
symmetry is then broken.

A random matrix formula for the time dependence of the decay
of an open quantum system has been derived by
Savin and Sokolov \cite{Savi97} using supersymmetry techniques.  
For CUE the case of perfect
coupling is realized~\cite{02wsrep} (i.e.~one has $T=1$ in eq.~(11) of 
ref.~\cite{Savi97}) and the decay probability depends only on the
number of channels $q$ and the ratio
$\tau =t/T_H$ where $T_H$ is the Heisenberg time $T_H=2\pi \hbar /\Delta$:
\begin{eqnarray}
P^{(q)}(t)\!=\!{\textstyle \frac{1}{2}}\!\int_{-1}^1\!\!\!d u\!\int_1^\infty\!\! \!dv\;
\frac{v\!+\!u}{v\!-\!u}\;\delta\left(2\tau \!+\!
u\!-\!v\right)\left[\frac{1\!+\!u}{1\!+\!v}\right]^q\!\!.
\label{PtInt1}
\end{eqnarray}
The integral 
can be expressed in terms of hypergeometric functions as
\begin{eqnarray}
P^{(q)}(t)=(1+\tau)^{-q}\,f_{\lessgtr}(\tau) \quad \textrm{for}\ 
 \tau \,{\scriptstyle \lessgtr}\, 1\,,
\label{PtInt2}
\end{eqnarray}
\begin{eqnarray}
f_<(\tau)\!\!\!&=&\!\!\!\sum_{k=0}^q\!{q\choose k}(-\tau)^k
\,\Big[\,\frac{1+\tau}{k\!+\!1}\,_2F_1\Big(q,k\!+\!1;k\!+\!2;\frac{\tau}{1\!+\!\tau}\Big)
\nonumber \\ 
&&\qquad \qquad
-\frac{2\tau}{k+2}\,_2F_1\Big(q,k\!+\!2;k\!+\!3;\frac{\tau}{1\!+\!\tau}\Big)\,\Big]\,,\\[2mm]
f_>(\tau)\!\!\!&=&\!\!\!
\frac{1+1/\tau}{q+1}\,_2F_1\Big(q,1;q+2;\frac{1}{1+\tau}\Big)\nonumber \\
&&\qquad \qquad -\frac{2/\tau}{(q\!+\!1)(q\!+\!2)}\,_2F_1\Big(q,2;q\!+\!3;\frac{1}{1\!+\!\tau}\Big)\,.
\label{PtHyp2}
\end{eqnarray}
Asymptotically the decay is algebraic, 
\begin{eqnarray}
P^{(q)}(t)\rightarrow \tau^{-q}/(q+1) \quad \textrm{for}\ \tau \rightarrow 
\infty \,,\label{RMT-asy}
\end{eqnarray}
where the exponent is equal to the number of decay channels.

Quantum stabilization, i.e.~algebraic decay 
was already observed in previous studies for a purely harmonic driving
(see~\cite{01universal} and references therein)  
but with markedly different decay exponents. For the symmetry broken driving
 (\ref{Fomega}) (and therefore with proper CUE symmetry contrary to the harmonic case)
also good quantitative agreement  with the channel number $q$ is found
as we will show in the following.

\begin{figure}[b]
\twoimages[width=6.6cm, clip]{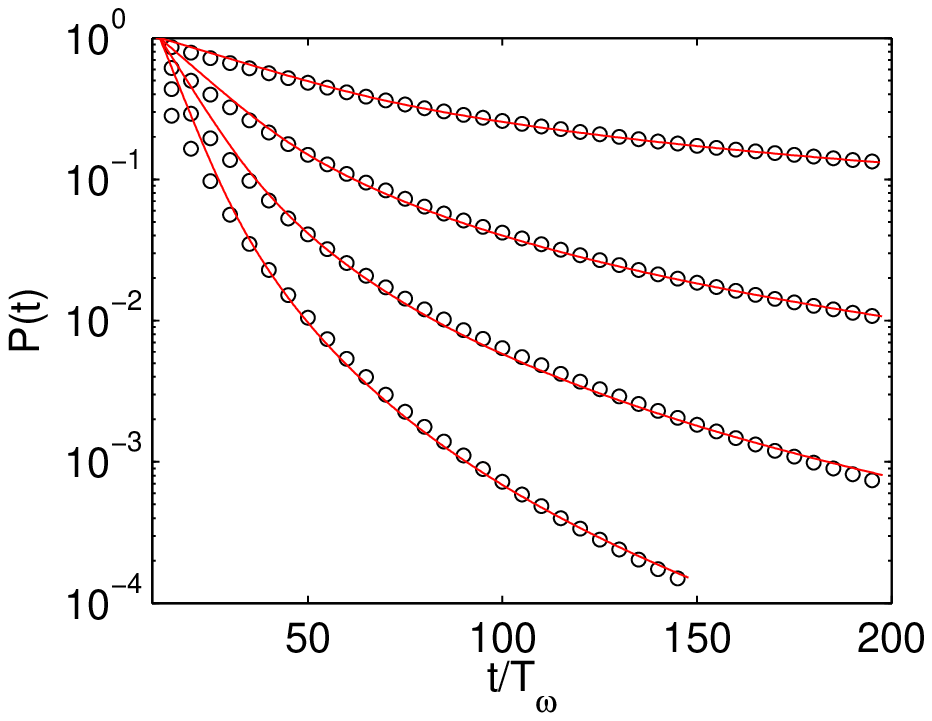}{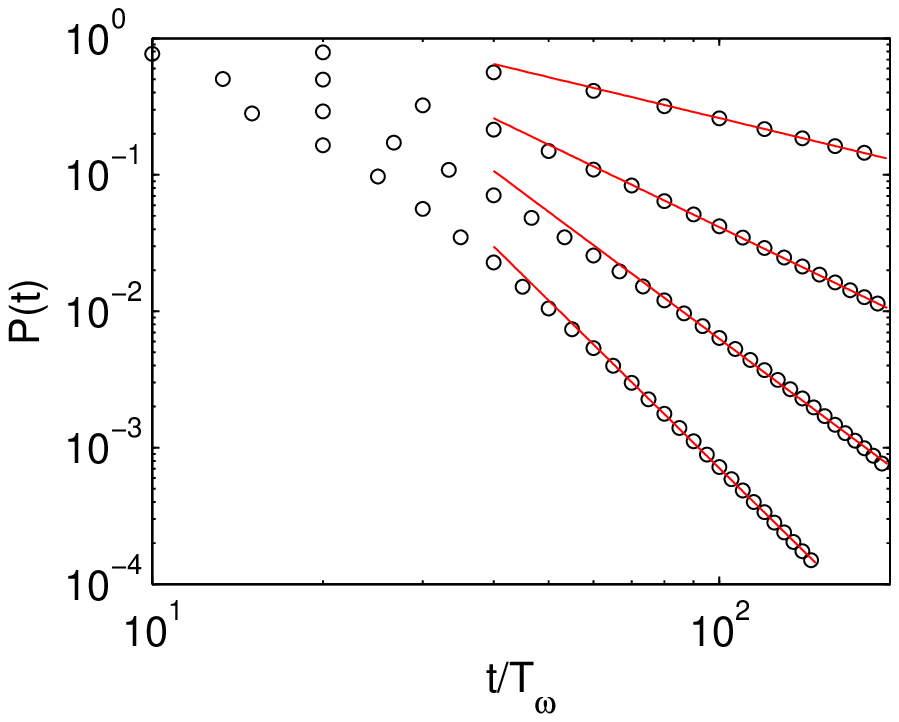}
\caption{Decay of the quantum survival probability $P(t)$ for rational 
frequency ratios
$\omega/\omega_B=1/q$ with  $q=1,\,2,\,3,\,4$ (open circles, from top to bottom) in comparison with
the results given by non-Hermitian random matrix 
theory for $q$ decay channels (solid lines). Right: double logarithmic plots
with asymptotic linear fits.}
\label{hyper}
\end{figure}

The system parameters are the same as in fig.~\ref{phsp-wave}, the
static force $F_0$ is tuned to give a rational ratio $\omega/\omega_B=r/q$
with $r=1$. 
A minimum uncertainty wavepacket with momentum width $0.28$
centered at $p=0$ and $x=\pi$ is propagated in time and the survival
probability in the chaotic strip $p_1 <p< p_2$ is calculated from the 
momentum distribution,
\begin{equation}
P(t)=\int_{p_1}^{p_2}\!\! d p\;\left|\psi(p,t)\right|^2\,,
\label{survival}
\end{equation}
where $p_1=-5$ and $p_2=7$ are chosen based on fig.~\ref{phsp-wave}. 
Numerically, these survival probabilities are calculated up to a
time $t=200\,T_\omega$ for $q=1$, $2$, $3$ and $4$\,.

A fit of the long time behavior of the numerical data to the
asymptotics (\ref{RMT-asy}) confirms the algebraic decay predicted
by random matrix theory: Numerically the exponents are calculated
as $0.999$, $2.006$, $3.094$ and $4.087$ for $q=1,\,2,\,3$ and $4$
in precise agreement with these channel numbers
(compare the double logarithmic plot in fig.~\ref{hyper}). The corresponding
Heisenberg times are found to be $T_H/T_\omega=53.2,\, 36.5,\,  30.0$
and $25.3$ and decrease approximately as $\sim q^{-1/2}$.
For resonant driving (\ref{q-r}) with $r\ne 1$ the decay is also found to be
asymptotically algebraic, however with exponents differing from $q$ which
is presumably due to the fact that the quasienergy spectrum is
$r$-fold degenerate which introduces additional symmetries into the system.

Figure \ref{hyper} shows a logarithmic plot of the numerical results
for the survival probabilities (\ref{survival})
as a function of $t/T_\omega$ up to $t=200T_\omega$ for rational ratios
$\omega/\omega_B=1/q$ for $q=1$, $2$, $3$ and $4$.
Also shown in fig.~\ref{hyper} is
the time dependence of the random matrix model (\ref{PtInt2}) 
using the values of the Heisenberg 
time extracted by the asymptotic fit. In order to account for the initial
spreading of the wavepacket to equilibrate over the chaotic strip, the numerical 
data are shifted by $8\,T_\omega$ in each case. Good agreement is observed.
The small deviations between the numerical data and the random
matrix predictions (in particular for $q=3$) are due to the
influence of small stability islands in the chaotic strip.

\begin{figure}[t]
\onefigure[width=6.0cm, clip]{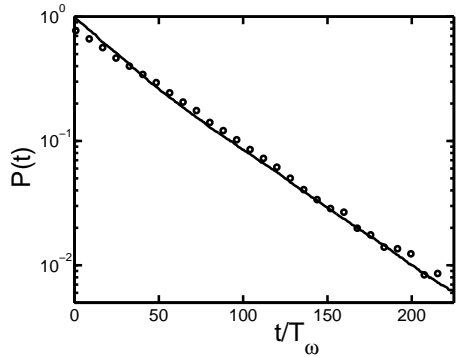}
\caption{Quantum survival probability $P(t)$ for irrational 
frequency ratios
$\omega/\omega_B=1/\sqrt{2}$ (full curve) in comparison with
the results from a corresponding classical ensemble (open circles).}
\label{irrational}
\end{figure}

For an irrational ratio $\omega/\omega_B$
the quantum survival probability decays exponentially, \,$P(t) = e^{-\nu t}$\,, 
as the classical distribution and even the exponent $\nu $ agrees
with the classical one. A comparison is given in fig.~\ref{irrational}. 
As shown by Puhlmann et al. \cite{Puhl05}, the random matrix predictions for
the survival probability can also be analyzed using a semiclassical
approach. A semiclassical treatment of the driven Wannier-Stark system
along these lines, however, deserves further studies.

\begin{figure}[t]
\onefigure[width=6.5cm, clip]{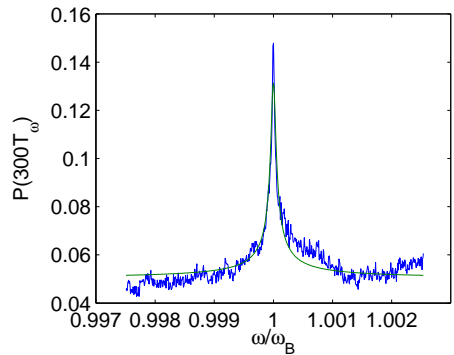}
\caption{Survival probability $P(t)$ as a function of 
$\omega/\omega_B$ in the vicinity of the resonance
$\omega=\omega_B$ for $t=300T_\omega$.}
\label{peaks}
\end{figure}

\section{Resonance profiles}
It is obvious that
the quantum survival probabilities $P(t)$ depend very sensitively on
the frequency ratio $\omega/\omega_B$.
Asymptotically, the decay is algebraic for rational and exponential
for irrational ratios and therefore a fractal-like spike structure
develops as already discussed in ref.~\cite{00transition2}.
As a typical example, fig.\,\ref{peaks} shows 
$P(t)$ as a function of 
$\omega/\omega_B$ in the vicinity of the resonance
$\omega=\omega_B$ for $t=300T_\omega$.
We observe a pronounced resonance peak which sharpens 
with increasing time. The maximum decays as $\sim t^{-1}$ and
the almost constant background decays exponentially. The
peak profiles can be well reproduced by the
functional form  
\,$P_{\rm fit}(t)=a/\sqrt{(\omega-\omega_B)^2+3(\Delta \omega)^2}+b$\,
also shown in the figure. The parameters $a$, $b$ and $\Delta \omega$
are functions of time.
 The time decay of the
width $\Delta \omega $ of the distribution was
found to be algebraic, $\Delta \omega \sim t^{-\gamma}$, 
with  $\gamma \approx 1.45$.
Interestingly, this is much faster than the  
Fourier behavior where two frequencies can only be
distinguished if their distance is at least of the
order the excitation time, i.e.~$\Delta \omega \sim t^{-1}$.
Similar observations of such `sub-Fourier' resonances have
been reported recently in an experimental study of bichromatically 
kicked cold atoms in an optical lattice~\cite{Szri02,Szri03}. 
This system also shows quantum chaotic
behavior and the mechanism of the sub-Fourier sharpening of these resonances 
could be explained theoretically~\cite{Lign04} based on the phenomenon of 
dynamical localization. 

\acknowledgments
Support from the Deutsche Forschungsgemeinschaft (GRK 792)
and from the Volkswagen-Stiftung is gratefully acknowledged.
S.M. acknowledges the support by a fellowship within the Postdoc-Programme 
of the the German Academic Exchange Service (DAAD).
We would also like to thank Jean-Claude Garreau and Dimitry Savin for 
discussions and Dominique Delande for providing a copy of 
ref.~\cite{Lign04} prior to publication.


\begin{thebibliography}{0}

\bibitem{Haak01}
\Name{Haake F.}  
\Book{Quantum Signatures of Chaos} 
\Publ{Springer, Berlin, Heidelberg, New York} \Year{2001}

\bibitem{Ishi95}
\Name{Ishio H. \and Burgd\"orfer J.} 
\REVIEW{Phys. Rev. B}{51}{1995}{R2013}

\bibitem{Wirt97}
\Name{Wirtz L.,  Tang J.-Z.\and Burgd{\"o}rfer J.}  
\REVIEW{Phys. Rev. B}{56}{1997}{7589}

\bibitem{Ishi00}
\Name{Ishio H.}  
\REVIEW{Phys. Rev. E}{62}{2000}{R3035}

\bibitem{Kott97Kott00}
\Name{Kottos T.\and Smilansky U.}  
\REVIEW{Phys. Rev. Lett.}{79}{1997}{4794};
\REVIEW{Phys. Rev. Lett.}{85}{2000}{968}

\bibitem{99delay99lifetime}
\Name{Gl{\"u}ck M., Kolovsky  A. R.  \and Korsch H. J.}  
\REVIEW{Phys. Rev. Lett.}{82}{1999}{1534}; 
\REVIEW{Phys. Rev. E}{60}{1999}{247}

\bibitem{02wsrep}
\Name{Gl{\"u}ck M., Kolovsky A. R. \and Korsch H.J.}  
\REVIEW{Phys. Rep.}{366}{2002}{103}

\bibitem{02deloc}
\Name{Korsch H. J.\ and Leyes W.}
\REVIEW{New J. Phys.}{4}{2002}{62}

\bibitem{04bloch1d}
\Name{Hartmann T., Keck F., Korsch H. J. \and S. Mossmann S.}
\REVIEW{New J. Phys.}{6}{2004}{2} 

\bibitem{Gasp98}
\Name{Gaspard P.}  \Book{Chaos, Scattering, and Statistical Mechanics}   
\Publ{Cambridge University Press, Cambridge} \Year{1998}

\bibitem{00transition2}
\Name{Gl{\"u}ck  M., Kolovsky A.R. \and Korsch H. J.}  
\REVIEW{Europhys. Lett.}{\bf 51}{2000}{255}

\bibitem{98ent}
\Name{Mirbach B. \and Korsch H. J.} 
\REVIEW{ Ann. Phys. (N.Y.)}{265}{1998}{80}

\bibitem{Zimm89}
\Name{Zimmermann T., K{\"o}ppel H. \and Cederbaum L. S.}  
\REVIEW{J. Chem. Phys.}{91}{1989}{3934}

\bibitem{Fyod96}
\Name{Fyodorov Y. V. \and Sommers H.-J.} 
\REVIEW{JETP Lett.}{63}{1996}{1026}; 
\REVIEW{J. Math. Phys.}{38}{1997}{1918}

\bibitem{Savi97}
\Name{Savin D. V. \and Sokolov V. V.}  
\REVIEW{Phys. Rev. E}{56}{1997}{R4911}

\bibitem{01universal}
\Name{Gl{\"u}ck  M., Kolovsky A.R. \and Korsch H. J.}  
\REVIEW{Physica E}{9}{2001}{478}

\bibitem{Puhl05}
\Name{Puhlmann, H., Schanz, H., Kottos, T. \and Geisel, T.}  
\REVIEW{Europhys. Lett.}{69}{2005}{313}

\bibitem{Szri02}
\Name{Szriftgiser P., Ringot J., Delande D. \and Garreau J. C.}  
\REVIEW{Phys. Rev. Lett.}{89}{2002}{224101}

\bibitem{Szri03}
\Name{Szriftgiser P., Lignier H., Ringot J., Garreau J. C. \and Delande D.}
 \REVIEW{Commun. Nonlin. Sci. Num. Simul.}{8}{2003}{301}

\bibitem{Lign04}
\Name{Lignier H., Garreau J.-C., Szriftgiser P. \and Delande D.} 
\REVIEW{Europhys. Lett.}{69}{2005}{327} 



\end{thebibliography}
\end{document}